# Extended sedimentation profiles in charged colloids: the gravitational length, entropy, and electrostatics

C.P. Royall[1], R. van Roij[2] and A. van Blaaderen[1]

[1]Soft Condensed Matter, Debye Institute, Utrecht University, PB 80000, 3508 TA Utrecht, The Netherlands,
[2]Institute for Theoretical Physics, Utrecht University, Leuvenlaan 4, 3584 CE Utrecht, The Netherlands.

tel: +31(0)30-2532952, fax: +31(0)30-2532706
www.colloid.nl, paddyroyall@netscape.net, r.vanroij@phys.uu.nl,
a.vanblaaderen@phys.uu.nl

Short title: Extended sedimentation profiles in charged colloids

**Abstract**

We have measured equilibrium sedimentation profiles in a colloidal model system with confocal microscopy. By tuning the interactions, we have determined the gravitational length in the limit of hard-sphere-like interactions, and using the same particles, tested a recent theory [R.van Roij, *J. Phys. Cond. Mat,* **15***,* S3569, (2003)], which predicts a significantly extended sedimentation profile in the case of charged colloids with long-ranged repulsions, due to a spontaneously formed macroscopic electric field. For the hard-sphere-like system we find that the gravitational length matches that expected. By tuning the buoyancy of the colloidal particles we have shown that a mean field hydrostatic equilibrium description even appears to hold in the case that the colloid volume fraction changes significantly on the length scale of the particle size. The extended sedimentation profiles of the colloids with long-ranged repulsions are well-described by theory. Surprisingly, the theory even seems to hold at concentrations where interactions between the colloids, which are not modeled explicitly, play a considerable role.

Classification numbers: PACS: 82.70.Dd, 05.20.Jj, 61.20.Gy

## 1. Introduction

A fundamental equilibrium property of colloidal suspensions in a gravitational field is the sedimentation profile. The competition between gravity, pulling the colloids down, and entropy, which promotes mixing, results in hydrostatic equilibrium, which for gradients on a scale larger than the size of the colloids is given by

$$\frac{dP(\rho(z))}{dz} = -mg\rho(z) \tag{1}$$

where $P(\rho)$ is the bulk osmotic pressure as a function of the colloid number density $\rho$, $z$ is the height, $m$ is the buoyant mass of each colloid and $g$ is the acceleration due to gravity. In the very dilute regime where the colloids may be considered as non-interacting (ideal gas limit), the equilibrium sedimentation profile is barometric, and the volume fraction $\eta=(\pi/6)\rho\sigma^{-3}$ of the colloid particles with diameter $\sigma$ decays exponentially with height, $\eta(z) = \eta_0 \exp(-z/L)$, where $\eta_0$ is the volume fraction at $z$=0. The decay constant is the gravitational length $L=k_BT/mg$, which was famously determined by Perrin in measurements of Boltzman's constant, $k_B$ (Perrin 1916). Here $T$ is the absolute temperature.

While the Barometric law is valid for dilute dispersions, at higher volume fractions, interactions between the colloidal particles become important, and the sedimentation profile, when integrated, yields the osmotic equation of state. The derivation of equation (1) assumes a local density approximation, which is only valid when the particle concentration can be assumed to be constant on length-scales several times larger than those over which the particles interact (Biben and Hansen 1993).

Charge-stabilised colloids present a multi component system, comprised of colloidal macroions and microscopic counter- and co- ions, all of which are immersed in a solvent. By adding salt to such a dispersion, the Coulombic repulsions between the colloids can be screened, so that the particles behave almost like a one-component hard-sphere system (Russel *et al* 1989). By integrating the sedimentation profile to obtain the osmotic pressure, equation (1), it was possible to obtain the hard-sphere equation of state (Piazza *et al* 1993), (Rutgers *et al* 1996). Intriguingly, Piazza *et al* (1993) reported that the gravitational length was up to twice as long as that expected on the basis of the particle diameter and the mass densities of the particles and the solvent. These workers suggested that this might be due to a failure of the one component fluid approach, that perhaps an electric field is created due to separation of colloids and counter-ions, which acts to pull the colloids up against gravity. Rutgers *et al* (1996) did not quantify the gravitational length in the top of the sediment, due to a weak signal at low colloid concentrations.

Since the gravitational length is so easily calculated with statistical mechanics, this discrepancy of a factor of two has caused considerable interest, and been investigated both theoretically (Biben *et al* 1993), (Biben and Hansen 1994), (Simonin 1995), (Loewen 1998), (Van Roij 2003), and with computer simulation (Hyninnen *et al* 2003). In the case of charge-stabilised colloids, the sedimentation profile is indeed predicted to be extended and for high colloid charges, this extension can be very considerable indeed, due to the separation of colloids and counter-ions suggested by Piazza *et al* (Piazza 1993). Nevertheless, in the limit of sufficient dilution (which may be such that $\eta \ll 10^{-4}$ for highly charged colloids at low ionic strength, but only $\eta \ll 10^{-1}$ for hard-sphere like systems) the barometric law is expected to be recovered.

Theories which explicitly consider the microscopic co- and counter-ions predict a separation of colloids and counter-ions, with more counter-ions at the top of the container, and more colloids at the bottom (Biben and Hansen 1994), (Van Roij 2003). This charge separation results in a macroscopic electric field, such that the whole system may be regarded as a condenser. In this electric field, the colloids are pulled upwards against gravity, to form an extended sedimentation profile. Upon adding salt, the co-ions can restore charge neutrality at all heights, and without the electric field, the sedimentation profile is barometric.

The Donnan equilibrium between an ion reservoir and a suspension of charged colloids is given by

$$\rho_{\pm} = \rho_i \exp\left(\mp \frac{e\psi}{k_B T}\right),$$

(2)

where $\rho_{\pm}$ is the cation (+) and anion (-) density in the colloid suspension, $2\rho_i$ is the number density of small cations and anions in the reservoir, $e$ is the electronic charge and $\psi$ is the Donnan potential. Imposing the charge neutrality condition $Q\rho = \rho_- - \rho_+$ in a suspension where charge on the colloids is $Qe$, yields within an ideal-gas assumption for the osmotic pressure $P = k_B T (\rho + \rho_+ + \rho_- - 2\rho_i)$. One rewrites this in dimensionless form as

$$\frac{PQ}{2\rho_i k_B T} = y + Q\left(\sqrt{1+y^2} - 1\right),$$

(3)

where $y=Q\rho/2\rho_i$ is a scaled colloid concentration defined by

$$y = 24\eta \frac{Q\lambda_B}{\sigma} \frac{1}{(\kappa\sigma)^2} \tag{4}$$

Here the Bjerrum length is denoted by $\lambda_B=e^2/(\varepsilon_S k_B T)$, where $\varepsilon_S$ is the permittivity of the solvent. The inverse Debye screening length of the ion reservoir is given by $\kappa = \sqrt{8\pi\lambda_B\rho_i}$ for monovalent ions. By combining the osmotic pressure as a function of colloid concentration (equation (3)) with the hydrostatic equilibrium (equation (1)), Van Roij (2003) showed that the sedimentation profile of charged (non-interacting) colloids does not necessarily give the Barometric distribution even though the charged species interact only with a mean field, not with one another explicitly. While the combined equations (1) and (3) must be solved numerically in order to obtain the exact density profile, it turns out to be possible to express the density profile analytically in three separate regimes, provided $Q>>1$,

$$y(z) = \begin{cases} y_0^{(1)} \exp\left(-\dfrac{z}{L}\right) & y < Q^{-1} & (i) \\ y_0^{(2)} - \dfrac{z}{QL} & Q^{-1} < y < 1 & (ii) \\ y_0^{(3)} \exp\left(-\dfrac{z}{(Q+1)L}\right) & y > 1 & (iii) \end{cases} \tag{5}$$

where the gravitational length is denoted by $L=k_B T/mg$. In this theory, both colloids and small ions interact *only* through the mean-field Donnan potential, which varies linearly with $z$ in regimes (*ii*) and (*iii*), and therefore gives rise to a bulk electric field. A more detailed Poisson-Boltzmann calculation is required to identify the source of this field, which is found to be due to a charge separation between top and bottom of the suspension. Such a calculation is required to fit experimentally determined sedimentation profiles, since equations (1) and (3) are only valid for systems with local charge neutrality.

At low volume fractions, equation (5) yields the familiar barometric regime (*i*). For intermediate volume fractions, a linear regime is expected (*ii*), while at high volume fractions a slowly decaying exponential regime is predicted (*iii*). Extended profiles in charge-stabilised colloidal dispersions have been observed experimentally (Philipse and Koenderink 2002), and recently, Rasa and Philipse (2004) have investigated the predictions of equation (5), for colloids of small diameter (19 nm) and charge (Q~10).

We seek to demonstrate that the theory also holds in a very different regime. In the model system used here, the colloids are two orders of magnitude larger in size, and have a correspondingly smaller gravitational length, and higher charge. Furthermore we determine the gravitational length, to investigate whether our system also exhibits the discrepancy observed by Piazza *et al* (1993). This is important because at the high salt concentrations they used, all effects of the electric field are predicted to vanish by the Poisson-Boltzmann theory due to screening, so the increase in gravitational length is not yet explained by theory.

We use confocal microscopy to obtain the co-ordinates of the colloids and thus the colloid volume fraction as a function of height in a very direct way, simply by counting the particles as a function of height. Measuring the particle co-ordinates also allows us to investigate effective interactions between the colloids, since the interactions uniquely determine the radial distribution function (RDF) (Henderson 1974), which we can calculate from the particle coordinates. By comparing the experimental RDF with Monte Carlo (MC) simulation (Royall *et al* 2003), we have another measure for the values of colloid charge and ion density that result from the results from fitting the sedimentation profile with Poisson-Boltzmann theory. We also obtain a further independent estimate of the ion concentration by conductivity measurements of the colloidal suspension.

**2. Experimental**

We used a new model system, of poly-methyl methacrylate (PMMA) colloids in a solvent mixture of cis-decalin and cyclohexyl bromide (CHB). By closely matching both the refractive index and mass density of the particles (Yethiraj and Van Blaaderen 2003), the system is optimised for confocal microscopy. The particles are fluorescently labelled with rhodamine isothiocyanate (RITC) according to the procedure described by Bosma *et al* (2002).

The diameter of the colloids was 1.91 $\mu$m, determined by scanning electron microscopy (SEM), and 1.95 $\mu$m determined by static light scattering (SLS), which suggests some degree of swelling upon dispersion. The polydispersity was found to be 4.7% in the SEM measurement. When we calculate the buoyancy of the particles, it is the mass of PMMA that is important, so we calculated this using the SEM diameter. The relatively large size and correspondingly slow dynamics of these colloids are suitable for confocal microscopy. For a Debye screening length in the range of the particle diameter, a solvent of extremely low ionic strength is required, such as the CHB-cis decalin mixture, whose ionic strength can reach $10^{-9}$ M. The dispersions were confined to glass capillaries with inner dimensions of 0.1 x 1.0 mm (VitroCom) and sealed at each end with Norland Optical adhesive no 68 (Norland Optical Products Inc). The capillaries were laid flat, and imaged from below. This configuration allowed the entire sample to be scanned, which is important for samples which sediment upwards. Using capillaries also minimises the effects of thermal convection. The samples were stored in the dark, and left for a week to equilibrate. No change was found, up to a week later.

To determine the sedimentation profiles, we measured the colloid coordinates, and thus obtained the profile simply by counting the particles. Our method to find the coordinates has been reported previously (Royall *et al* 2003), and is briefly recapped here. A bright pixel in a 3D confocal image was identified as a trial colloid centre. The intensity of neighbouring pixels was used to determine the centroid of the object, and this centroid was then taken as the colloid coordinate. This resulted in an error of around $0.1\sigma$ in the position of each colloid, which does not matter for the sedimentation profile measurement. The scan is sufficiently fast that each particle was assumed to be fixed, although, since it takes up to 30 s to scan the whole 3D image, some diffusion of the top particles took place before the scan reached the bottom, however this does not matter as we seek only an average measure of the number of co-ordinates at each height. Typically, we sampled 32 3D images, which consisted of 256 *xy* frames of 256x256 pixels each. The pixel size was 195 nm for *xy* and 390 nm in the *z* direction. For the measurements of the radial distribution function (see below), it is important that all co-ordinates are sampled 'instantly'. In this case, we scanned sufficiently quickly (~2s) compared to the time required for the colloids to diffuse one diameter (~30s), that the particles could be assumed to be fixed. Here we reduced the size of the sampled region to typically 16 *xy* frames.

Since we used a mixture of solvents to density match the colloids, we could vary the buoyant mass of the particles by changing the solvent composition and hence the solvent density. This allowed us to determine the density of the PMMA particles, which is not known *a priori*, but is expected to differ from the bulk density of PMMA. To measure the gravitational length, we prepared samples with tetra butyl ammonium bromide salt (TBAB), which is known to screen the weakly charged colloids, and give a hard-sphere like behaviour (Yethiraj and Van Blaaderen 2003). All samples were prepared with 1% volume fraction of colloids, and the solvent compositions are listed in table (1). The samples were then termed T70 for 70% CHB by mass, and so on. The samples were prepared from a stock solution of 160 $\mu$M TBAB in CHB, and diluted with cis-decalin. The resulting molar strengths are listed in table (1), although the TBAB dissociates to less than 1%, so the resulting ion concentration is of the order of $\mu$M. We prepared six samples with TBAB salt and varied the solvent composition between 70% and 100% CHB by mass.

Evaluation of the gravitational length also requires the density of the solvent mixture. This we measured at 20.0 °C, with an Anton Paar density meter. We fitted our measurements to the following polynomial

$$\rho_s = 0.904 + 0.576 f - 0.585 f^2 + 0.441 f^3$$

(6)

where $\rho_s$ is the density of the solvent in gcm$^{-3}$, and $f$ is the mass fraction of CHB in the solvent mixture.

## 3. Results and discussion

*3.1. Determination of the gravitational length with screened charged colloids*

To check that the TBAB salt does indeed screen the particles effectively, we investigated the colloid-colloid interactions via the radial distribution function (RDF). Our method here has been described previously (Royall *et al* 2003). Briefly, we determined the particle co-ordinates as described above, and binned the interparticle distances into a histogram to yield the RDF, normalising appropriately to account for the finite size of the microscope image. We then fitted the experimental RDF with one obtained for a known interaction, from Monte-Carlo (MC) simulation in the canonical ensemble (Frenkel and Smit 2002). In the simulation, the effective colloid-colloid interactions are described by a Hard-Core Yukawa interaction. We neglect the Van der Waals attractions because the colloids and solvent are refractive index-matched. By a linearisation of Poisson-Boltzmann theory (Verwey and Overbeek 1949), the interaction potential $u(r)$ may be related to the colloid charge and Debye screening length by

$$\beta u(r) = \begin{cases} \infty & r \leq \sigma \\ \dfrac{Q^2}{(1+\kappa\sigma/2)^2} \dfrac{\lambda_B}{\sigma} \dfrac{\exp(-\kappa(r-\sigma))}{r/\sigma} & r > \sigma \end{cases} \quad (7)$$

Since this analysis is only valid for a homogenous fluid, we did not analyse data sets with concentration gradients, and we only determined the RDF for the sample with the best density matching, T75 (see table 1), and for this sample we only took a thin slice around 8 $\mu$m from which we sampled the coordinates. As the RDF plots in figure 1 show, the colloids are not totally screened by the salt, and show some residual repulsion in addition to the hard-core interactions. However, reference to equation (5), with the fitted values of $Q$ of around $200\pm50$ and Debye length of $100\pm20$ nm suggests that, although the Coulombic repulsions are not totally screened, the scaled colloid concentration, $y$, is at least an order of magnitude below $1/Q$ and so the sample should be in regime (*i*) of equation (3) and hence should obey the barometric law. For the short screening length, the inaccuracy in finding the colloid coordinates is important, as it distorts the RDF, particularly at higher volume fractions. In the $\eta$=0.12 case, we needed to add an 'error' of $0.1\sigma$ to the simulation co-ordinates to obtain agreement with the experimental RDF.

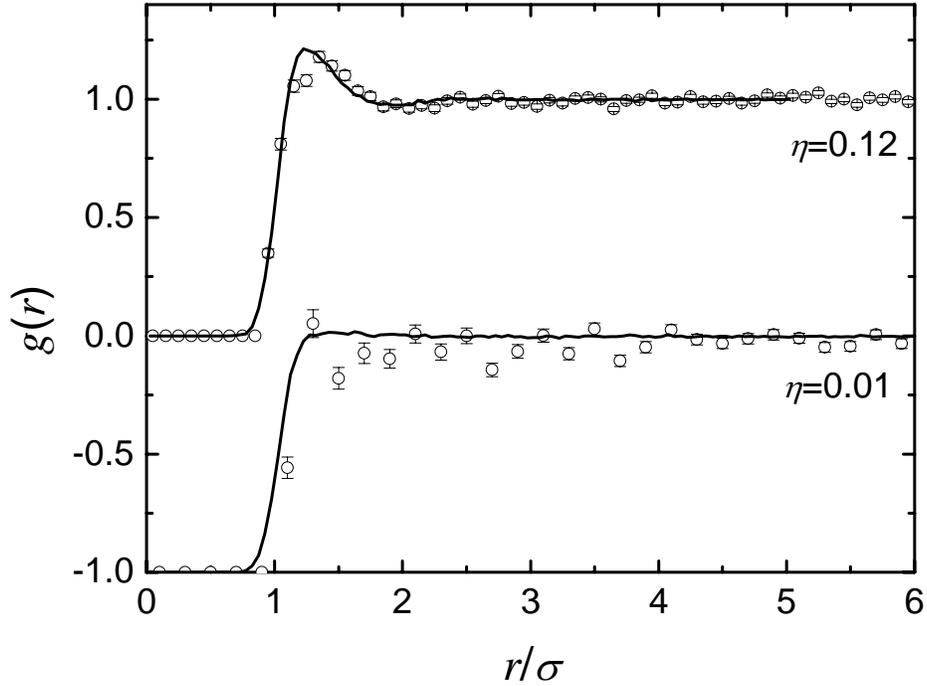

*Figure 1. The radial distribution function for the T75 sample. Circles are experimental data, black lines are results from MC simulation using equation (7), with parameters of $Q=200\pm50$, $\kappa\sigma=20\pm4$. The plot for $\eta=0.01$ is shifted downwards for clarity.*

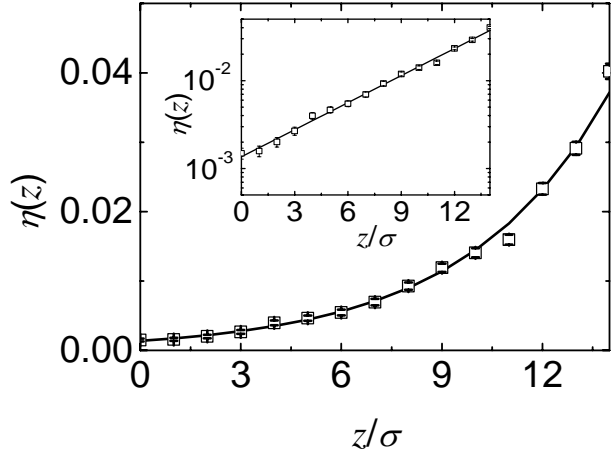

*Figure 2. Sedimentation profile of the T80 sample. Inset shows a log plot of the same data, revealing the expected exponential decay. The gravitational length is found from fitting the profile with the lines shown, for volume fractions less than 3%.*

| sample | mass fraction CHB | TBAB ($\mu$M) | $L/\sigma$ |
| --- | --- | --- | --- |
| T70 | 0.701 | 128 | 3.23±0.11 |
| T75 | 0.745 | 134 | 13.1±0.3 |
| T80 | 0.799 | 140 | -3.61±0.10 |
| T85 | 0.802 | 141 | -3.32±0.05 |
| T90 | 0.898 | 151 | -0.98±0.02 |
| T100 | 1 | 160 | -0.55±0.02 |

*Table 1. Sample parameters for the 'hard-sphere' like systems. The gravitational length L is found from fitting plots such as figure (2), and the error shown is due to the uncertainties from the fitting proceedure only.*

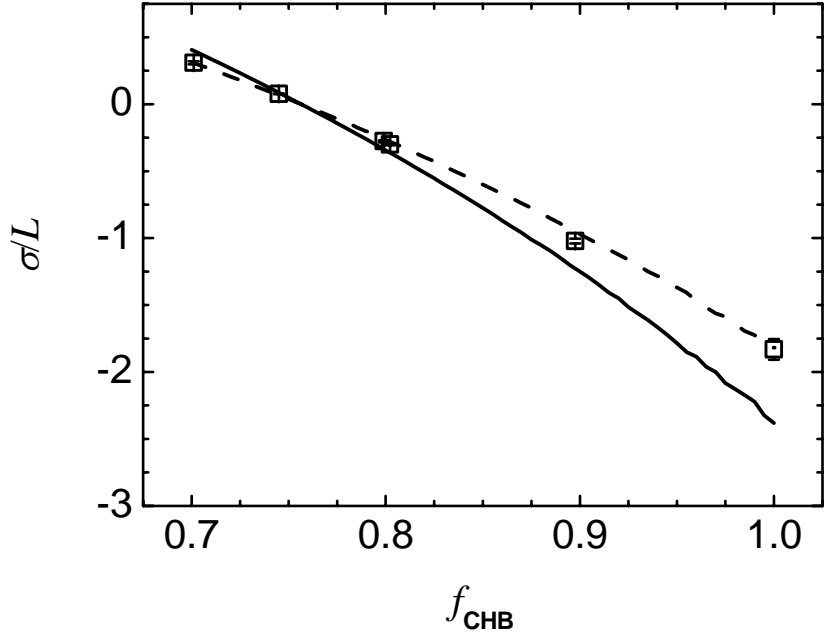

*Figure 3. Inverse gravitational length as a function of CHB mass fraction for samples with added salt. The solid line assumes a colloid diameter of 1.91 μm as measured by SEM, while the dotted line which gives better agreement assumes an increase in the gravitational length of 11%.*

Typical results for the sedimentation profile are shown in figure 2. The exponential behaviour is clear, and using the gradient of the straight line obtained from the linear-log plot (inset), we calculated the gravitational length. We list all six gravitational lengths in table 1. In all samples, we fitted the slope for volume fractions less than 3% volume fraction, as at higher concentrations, interactions between the colloids cannot be neglected. Figure 3 shows the predicted and measured inverse gravitational length, comparing the experimental values with those calculated, assuming a PMMA density of 1.196 g cm$^{-3}$ throughout. Note that the density of PMMA is fixed between the bounds of 1.191 gcm$^{-3}$ and 1.215 gcm$^{-3}$, because for the T75 sample, the colloids sink, while for the T80 sample they float. Although there is some discrepancy, such that the gravitational length is around 11±1% larger than the calculated value, the difference is much less than a factor of two, as found by Piazza *et al* (1993). Nonetheless, we investigated possible causes for this discrepancy. In particular, the system has some polydispersity, so we may expect some size segregation in the gravitational field. To quantify this effect, we conducted Monte Carlo simulations of a dilute, hard-sphere system with a polydispersity of 4.7% in a gravitational field. We determined the gravitational length in the same way as in figure 2, by fitting the exponential decay, and found an increase in the gravitational length of 8±2% due to the polydispersity. Hence we believe that the increase in gravitational length measured is largely accounted for by the polydispersity of the PMMA colloids. A further slight increase in the gravitational length may be attributed to the fact that the effective mass of the particles includes a layer of bound solvent, possibly of the same lengthscale as the stabliser layer, 11nm in pure cis decalin (Bryant *et al* 1999). The original observation by Piazza *et al* (1993) of a larger gravitational length therefore remains unexplained.

Our results also suggest that mean-field hydrostatic equilibrium (equation (1)) applies even when the volume fraction varies significantly on the colloid lengthscale: For the T100 sample, the gravitational length is around half a diameter, yet no change in sedimentation profile behaviour is seen. However in the dilute regime where these

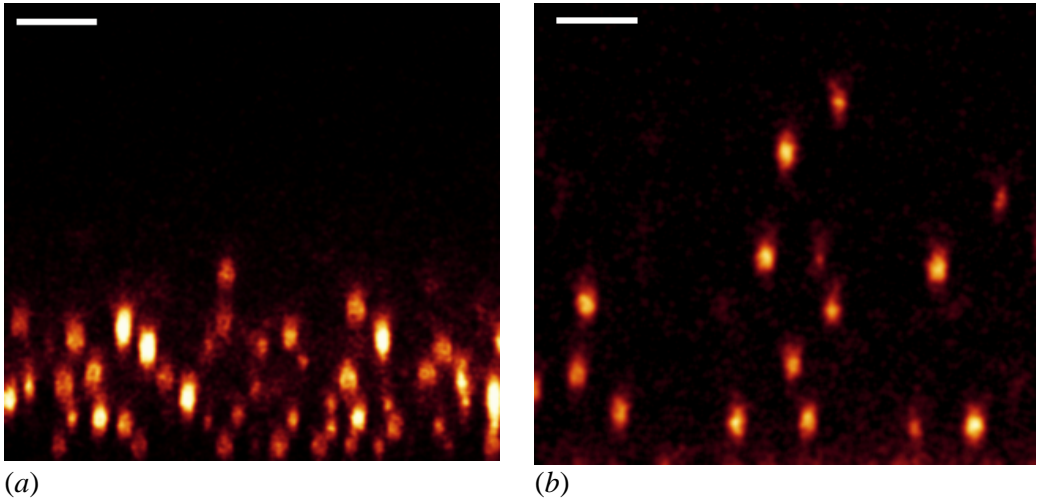

(*a*)  (*b*)

*Figure 4. xz 'slices' from confocal microscope images showing the T70 (a) and C70 (b) samples. Both samples have the same degree of density mismatch, the only difference is that the T70 system has the TBAB salt added to screen the electric field. Bars=10μm.*

measurements were taken the local density approximation assumed by equation (1) holds, so good agreement even in the case that $L<\sigma$ is to be expected.

*3.2. Charged colloids: extended sedimentation profile*

To investigate the extended sedimentation profile in the case of colloids with long-ranged interactions, we prepared four samples with no added salt. The details are listed in table 2. Again we varied the CHB mass fraction from 70% to 100% and

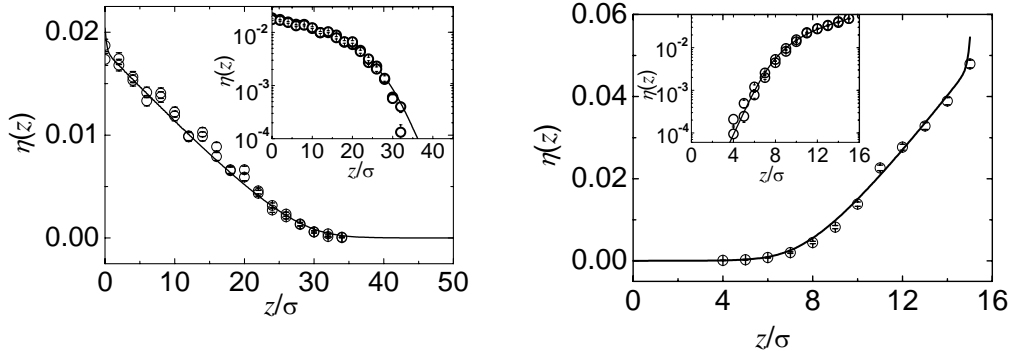

(*a*)                                (*b*)

*Figure 5. Sedimentation profiles for C70 (a) and C90 (b). Insets are logarithmic plots of the same data. Circles are experimental data, while lines are the solutions to the Poisson-Boltzmann theory for the parameters listed in table (2). The crossover from barometric (i) to linear (ii) behaviour is clear, and is consistent the theory. It is also possible to obtain the gravitational length from these plots, which is consistent with that obtained from when salt is added. Note that the different direction of the slope reflects the fact that in the C70 (left) sample, the colloids sink, while in C90 (right) they float.*

termed these samples C70 for 70% CHB, and so on. The experimental procedure was as described above. The extended sedimentation profile in the case of no added salt is clearly illustrated in figure 4, which shows an *xz* 'slice' of a 3d confocal microscope image. The samples shown in figure 4 are the same, except that in figure 4(*a*) TBAB salt is added and figure 4(*b*) it is not. Typical sedimentation profiles are shown in figure 5 for C70 (*a*), and C90 (*b*). The different signs of the slope reveal the fact that for C70, the colloids sink, while for C90, they float. Both plots reveal region (*i*) and

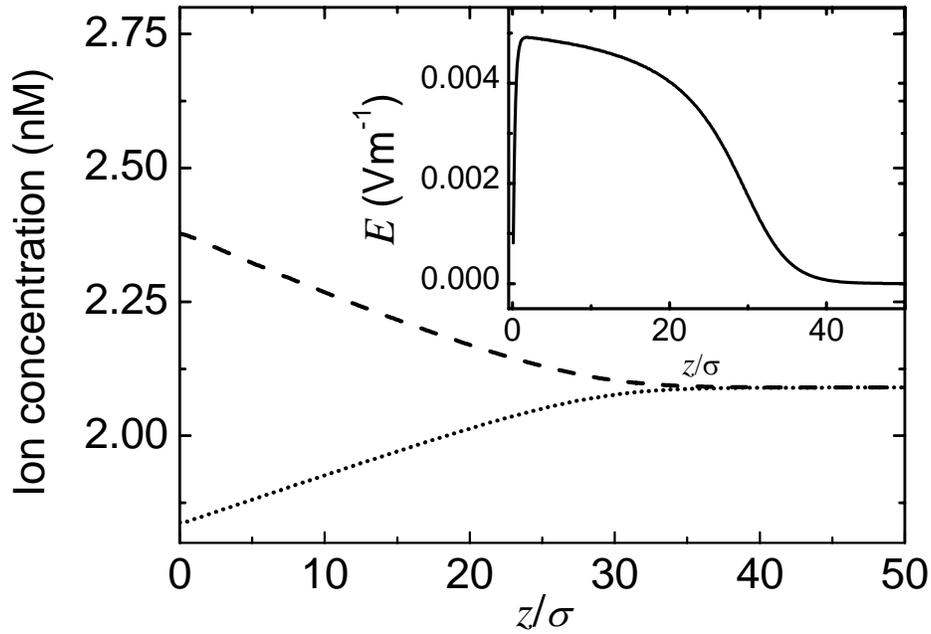

*Figure 6. Ion concentration profiles, for positive (dotted) and negative (dashed) ions for the C70 sample. Inset is the inhomogenous electric field which causes the extended sedimentation profiles.*

(*ii*) as suggested by fitting the sedimentation profiles with Poisson-Boltzmann theory. By fitting the profiles we arrive at a value for the ratio of the colloid charge and ion concentration (table 2). The quality of the fit is only sensitive to the ratio $Q/\kappa\sigma$, such that both $Q$ and $\kappa\sigma$ can be varied by a factor of two with little change in the result. Knowing one or both independently allows us in the first case to evaluate the other, and, in the second case, to test equations (1) and (3). Evaluating equation (3) with equation (1) gives the three regimes simultaneously, and we find that both the crossover from the Barometric regime (*i*) to the linear regime (*ii*) and the slope of the linear regime are consistent with the theory. We do not find the third regime as we optimised our study towards regimes (*i*) and (*ii*) where the theory is expected to be more accurate.

We also calculated the ion concentrations for the C70 sample (figure 6), which are smooth and almost constant. This is consistent with the mechanism that the ion entropy (which is maximised for homogeneous profiles) is the driving force for the electric field which raises the colloids to greater heights. Figure 6 also shows that the ionic strength hardly changes throughout the sedimentation profile, consistent with equation (2) in the regime $y<1$, such that $e\psi/kT<1$. The inset of figure 6 shows the electric field, also calculated according to the Poisson-Boltzmann theory. Since the electric field is related to the charge density by

$$\frac{dE}{dz} = -\frac{4\pi}{\varepsilon_S}\left(Q\rho + \rho^+ - \rho^-\right) \tag{8}$$

from the slope in the figure 6, inset, we see that the positive charge is localised at the bottom of the sedimentation profile, and balanced by a more widely distributed negative charge higher up in the sediment, for $z>\sigma$.

We obtained an independent measure for the ion concentration from conductivity measurements, but to convert a conductivity measurement to an ionic strength, we assume Walden's rule (Atkins 1994), and that the limiting molar conductance of the ions is equal to that of Bromide, 78.1 cm$^2$ S mol$^{-1}$. This is a significant assumption, and therefore the results obtained for the ionic strength should be treated with some caution (table 2).

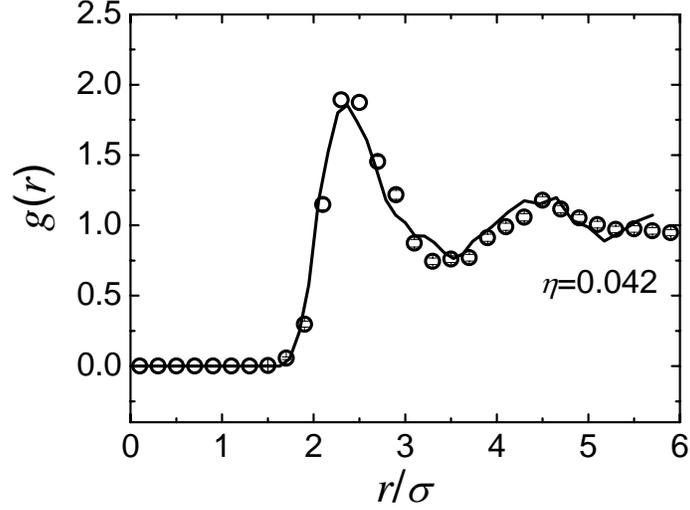

*Figure 7*. Radial distribution function determined experimentally (circles) for the C80 sample, and fitted from Primitive model Monte-Carlo simulation, for $Q=200\pm40$, and an ion reservoir with $\kappa\sigma=1.5\pm0.3$, black line.

| sample | mass fraction CHB | $Q/\kappa\sigma$ | $\lambda_B$ (nm) | Conductivity (pScm$^{-1}$) | $\kappa\sigma$ | $Q$ |
|---|---|---|---|---|---|---|
| C70 | 0.701 | 64.4 | 10.4 | 78.8 | 1.2 | 78 |
| C80 | 0.797 | 89.7 | 9.15 | 243 | 2.0 | 174 |
| C90 | 0.899 | 40.3 | 7.99 | 660 | 2.9 | 119 |
| C100 | 1 | 54.2 | 7.10 | 1466 | 4.0 | 219 |

*Table 2*. Sample parameters for the unscreened systems with no added salt. The ratio $Q/\kappa\sigma$ is determined by fitting the measured sedimentation profiles such as figure (5) with equations (1) and (3). $\kappa\sigma$ was determined from conductivity measurements, and used to evaluate the colloid charge $Q$, using the values for the ratio $Q/\kappa\sigma$.

We also have another measure for the ratio of $Q/\kappa\sigma$ from the RDF analysis as before. Now at these relatively low colloid charges and large screening lengths, the DLVO model (equation 7) can under-estimate the structure (Loewen 1993), so fitting the RDF leads to an over-estimation for the colloid charge. Consequently, instead of the Yukawa interaction, we used the primitive model, which explicitly considers all the small ions in the system, and treats the solvent as a dielectric medium. Briefly, we use an unrestricted primitive model, where the particles interact via a Coulomb potential with a hard core:

$$\beta u_{ij}(r) = \begin{cases} \infty & r_{ij} < \frac{1}{2}(\sigma_i + \sigma_j) \\ \frac{q_i q_j \lambda_B}{r_{ij}} & r_{ij} \geq \frac{1}{2}(\sigma_i + \sigma_j) \end{cases} \quad (9)$$

where $i$ and $j$ are the interacting species, and may be colloids (charge $+Qe$) or monovalent co- or counter-ions. The model is implemented on a lattice, where the lattice spacing is chosen such that each colloid diameter is divided into 19 lattice sites. The simulation method is based on that used by Panagiotopoulos and Kumar (1999). Again, we vary the ion density and colloid charge such that the simulation agrees with experiment. We measured the RDF of the C80 sample only, as before, this sample was chosen due to its relatively good density-matching, such that the sampled region was quite homogeneous.

The results are shown in figure 7, for a value of Q=200±40 and $\kappa\sigma$=1.5±0.3 in the ion reservoir. This gave a ratio of 133 for $Q/\kappa\sigma$, and the value of 90 obtained from fitting the sedimentation profile with Poisson-Boltzmann theory lies within the error bounds. Thus two very different methods to determine the ratio $Q/\kappa\sigma$ yield similar values. Furthermore, we obtained a value of $\kappa\sigma$=2.0 from conductivity measurements, which compares reasonably with that from the simulation data. For the other samples, we use conductivity to estimate the Debye length to calculate the charge listed in table 2. There is a trend to reduce the charge with the addition of a greater quantity of cis-decalin: this may be related to the change in dielectric constant of the solvent mixture, which reduces from 7.9 to 5.37 over the range sample (table 2). These charges of order Q~100 are comparable to those measured by electrophoresis on similar samples.

As above, we note that hydrostatic equilibrium appears to hold, even in the case of pure CHB (C100), where the colloid volume fraction changes on the lengthscale of the particle size. Another, conclusion is that the theory gives good agreement even when the colloids interact as suggested by the RDF in (figure 7). We measured the volume fraction of the fluid-body centred cubic crystal transition, and found it to be around 7.5% volume fraction. While there is some change in the colloid-colloid interactions due to the varying solvent composition, we see that figure 5(*b*) in particular shows good agreement between the theory and experiment up to 4% volume fraction. This agreement is surprising when we recall that the only interactions between the various charged species considered by the theory are smeared (mean-field) electrostatics and gravity. Note that similar conclusions may be reached by considering the profiles shown in the simulation data of Hynninen *et al* (2003).

## 4. Conclusions

In conclusion, we have measured the gravitational length for a colloidal model system with hard-sphere-like behaviour, and found it to be close to that expected. By calculating the effects of the relatively small polydispersity of our system, we find almost perfect agreement with theory and so we do not find the discrepancy in gravitational length noted in the literature (Piazza *et al* 1993). We have also shown that hydrostatic equilibrium appears to hold locally, even when the colloid volume fraction changes on the lengthscale of the particle size.

By tuning the interactions, we can also test a recent Poisson-Boltzmann theory for charge-stabilised colloids. In agreement with the theory, we find an exponential and linear regime in the sedimentation profile. Both regimes are in quantitative agreement with the theory. Furthermore, comparison with the theory yields parameters for the colloid charge and ion density. These were determined independently, by analysing the RDF, and found to be in reasonable agreement. Furthermore, we have a *second independent* measure for the ionic strength from conductivity, which also gives good agreement. Hence we conclude that the theory outlined in equations (3-5) gives clear quantitative agreement with the model system used here, even when there are very considerable interactions between the colloids.

Acknowledgements

It is a pleasure to thank Antti-Pekka Hynninen for providing the primitive model code, modified from that of Athanassios Panagiotopoulos and Marjolein Dijkstra for generous provision of CPU time. Albert Philipse, Mircea Rasa and Hartmut Löwen are thanked for stimulating discussions, and Didi Derks and Yu-Ling Wu for help with particle synthesis, and Mirjam Leunissen for help with conductivity measurements. This work is part of the research program of the "Stichting voor Fundamenteel Onderzoek der Materie (FOM)", which is financially supported by the "Nederlandse Organisatie voor Wetenschappelijk Onderzoek (NWO)".